\documentclass{epl}
\usepackage{amsmath}

\title{A ferromagnet with a glass transition} 
\shorttitle{}

\author{
S. Franz~\inst{1}\thanks{E-mail:
\email{franz@ictp.trieste.it}}
\and M. M\'ezard~\inst{2}\thanks{E-mail:
\email{mezard@ipno.in2p3.fr}}
\and F. Ricci-Tersenghi~\inst{1}\thanks{E-mail:
\email{federico.ricci@ictp.trieste.it}}
\and M. Weigt~\inst{3}\thanks{E-mail:
\email{weigt@theorie.physik.uni-goettingen.de}}
\and R. Zecchina~\inst{1}\thanks{E-mail:
\email{zecchina@ictp.trieste.it}}
}

\institute{
\inst{1} The Abdus Salam International Centre for
Theoretical Physics, Condensed Matter Group -
Strada Costiera 11, P.O. Box 586, I-34100 Trieste, Italy\\
\inst{2} LPTMS, Universit\'e de Paris Sud -
Bat 100, F-91405 Orsay, France\\
\inst{3} Institute for Theoretical Physics, University of
G\"ottingen - Bunsenstr. 9, D-37073 G\"ottingen, Germany
}

\pacs{05.70.Jk}{Critical point phenomena}
\pacs{64.70.Pf}{Glass transitions}
\pacs{75.10.Nr}{Spin-glass and other random models}

\begin{document}

\maketitle

\begin{abstract}
We introduce a finite-connectivity ferromagnetic model with a
three-spin interaction which has a crystalline (ferromagnetic) phase
as well as a glass phase. The model is not frustrated, it has a
ferromagnetic equilibrium phase at low temperature which is not
reached dynamically in a quench from the high-temperature
phase. Instead it shows a glass transition which can be studied in
detail by a one step replica-symmetry broken calculation. This spin
model exhibits the main properties of the structural glass transition
at a solvable mean-field level.
\end{abstract}

Over the last decades there has been a growing interest in
understanding the structural glass transition in complex materials.  A
major theoretical role has been played by mean-field theories based on
fully connected p-spin glass
models~\cite{MPV,Thirumalai,YoungBook}. In such models, it was
possible to identify a purely dynamical transition which accounts for
the off-equilibrium behaviour of the models, and which is believed to
present some analogies to the structural glass transition in realistic
systems~\cite{Gotze,CriHoSom,KuCu,BouCu}.

Although this analogy is very fruitful, one must keep in mind that the
mean-field spin glass models which have been considered are rather
remote from the structural glasses. Let us mention just the three most
obvious differences: they have no crystalline state, their Hamiltonian
has quenched-in disorder, in the form of an extensive number of
quenched coupling constants, and these couplings are of infinite
range. A lot of work has already been devoted to cure these defects,
in order to get a much closer analogy between some lattice spin models
and structural glasses. This should allow to clarify some basic issues
as for instance the presence of heterogeneities or the role of
geometrical frustration.  A few years ago, it has been realized that
the existence of quenched disorder is not a necessary ingredient for a
system to have a spin glass phase: several mean-field spin models were
constructed which display a discontinuous spin glass transition
typical of the p-spin glasses~\cite{berna}. More recently, numerical
simulations of some unfrustrated 3-dimensional spin problems, with
purely ferromagnetic multi-spin interactions, have shown a glassy
behaviour which persists for very long times~\cite{p-spin-3d}. A
similar behaviour was found also analytically in 2
dimensions~\cite{p-spin-2d}.

In this paper, we introduce a similar ferromagnetic model having
multi-spin interactions and finite connectivity. The model is defined
on a Bethe hyper-lattice (or Husimi tree, the analog of a Bethe
lattice for systems with plaquette interactions), where it can be
solved completely using the replica/cavity method.  This model thus
provides the correct mean-field (Bethe-)approximation for these
ferromagnetic multi-spin systems. As in the usual Bethe lattice, it is
a finite-connectivity system (each spin interacts with a finite number
of neighbours), but has a locally tree-like structure which makes it
solvable at the mean-field level.  The spin-glass version of the
model, with random boundary conditions, was considered first
in~\cite{riki}.  Our solution shows a phase diagram qualitatively
similar to that of usual p-spin models, with a static and a dynamic
glass transitions appearing at two different temperatures. On top of
the glass phase, it also displays a purely ferromagnetic phase which
is the analog of the crystalline phase in supercooled liquids.

The model is defined by the Hamiltonian
\begin{equation}
H = - \sum_{[i,j,k]\in E} J_{ijk} S_i S_j S_k
\label{eq:h}
\end{equation}
where $S_i=\pm 1$ ($i=1,...,N$) are Ising spins.  The model is
ferromagnetic: we take $J_{ijk}=1$ for all the triples $[i,j,k]$
belonging to the set $E$ of hyper-edges (plaquettes).  So the system
is not frustrated and the configuration $S_i=1$ ($i=1,..,N$) is a
ground state for every possible choice of $E$ and satisfies
simultaneously all interactions. The plaquettes are chosen randomly,
according to the two following ensembles:
\begin{itemize}
\item In the first ensemble, the only constraint is that each spin
belongs exactly to $k+1$ plaquettes.  The set of plaquettes builds up
a hyper-graph which is locally a Husimi tree of fixed connectivity
$k+1$.  This structure of hyper-edges eventually loops back onto
itself, but the typical length of the loops scales like $\ln(N)$. This
hyper-graph was used analogously in frustrated spin models studied in
the context of spin glasses~\cite{diluted}. In particular one should
notice that its structure is not disordered on any finite length
scale: disorder comes in only through the loops which have diverging
length scale and so the system is only very weakly disordered.
Basically this construction amounts to having a Cayley hyper-tree with
random hyper-edges closing the boundary.
\item Our results can be easily generalized to hyper-trees with
fluctuating connectivities.  Such models, with Poisson-distributed
connectivities, were introduced recently in~\cite{BaZe,RiWeZe} as a
simple case of an optimization problem. As in the case of
ferromagnetic systems with multi-spin interactions in finite
dimensions~\cite{p-spin-3d,p-spin-2d}, it was observed that purely
ferromagnetic interactions may lead to an effective dynamical
frustration. This was shown in~\cite{RiWeZe} by comparing lowest
metastable states in a ferromagnetic finite-connectivity 3-spin model
with the ground states of the corresponding frustrated spin-glass
version ($J_{ijk}=\pm1$ randomly), which were found to share exactly
the same statistical features. As also seen in~\cite{BaZe} by
analyzing the off-equilibrium low temperature dynamics of both models,
the ferromagnetic system is not able to equilibrate due to the
entropic dominance of metastable states.
\end{itemize}
The results can also be generalized to interactions between clusters
of $p$ spins ($p\geq 3$), also with spins belonging to clusters but
interacting pairwise, which can be viewed as a representation of
geometrical frustration.

At first we concentrate on an analytical approach to the equilibrium
behaviour based on the replica trick~\cite{MPV}. The results will be
compared later on with data obtained form Monte Carlo simulations.
The free-energy density can be read off from the $n\to 0$ limit of the
$n$-fold replicated free energy at temperature $T=\beta^{-1}$. The
formulas are written here for a general hyper-tree with $k+1$
hyper-edges per site, we shall then discuss in more details the case
$k=3$.  This is the lowest value of $k$ for which ferro-magnetic and
glassy behavior are present, for $k=2$ the system is paramagnetic at
all temperatures~\cite{riki}.

The natural order parameter in the replica approach is a probability
distribution $c(\vec\sigma)$ on the set $\vec\sigma\in\{\pm 1\}^n$,
which counts the fraction of sites $i$ having replicated spin
$S_i^a=\sigma^a$ ($a=1,...,n$).  For details on this approach see
Refs.~\cite{diluted,RiWeZe}. In terms of this order parameter, the
replicated free energy reads
\begin{equation}
-\beta f_n = \underset{c(\vec\sigma)}{\mbox{extr}}
\left[
        {k\!+\!1 \over 3} \ln 
        \left(
                \sum_{\vec\sigma_1,\vec\sigma_2,\vec\sigma_3}
                c(\vec\sigma_1)^k c(\vec\sigma_2)^k c(\vec\sigma_3)^k 
                e^{\beta\sum_{a=1}^n \sigma_1^a\sigma_2^a\sigma_3^a} 
        \right)
        -k\ln
        \left(
                \sum_{\vec\sigma} c(\vec\sigma)^{k\!+\!1} 
        \right)
\right]
\label{eq:fn}
\end{equation}
We notice that if we had considered spin-glass couplings $J_{ijk}=\pm
1$ with symmetric probability, we would have found $\cosh( \beta
\sum_{a=1}^n \sigma_1^a \sigma_2^a \sigma_3^a)$ instead of the
exponential in (\ref{eq:fn}).

One needs to find a distribution $c(\vec\sigma)$ which makes this free
energy stationary, which means that it must satisfy the saddle point
equation
\begin{equation}\label{eq:csigma}
c(\vec\sigma) \propto \sum_{\vec\sigma_2,\vec\sigma_3}
c(\vec\sigma_2)^k c(\vec\sigma_3)^k \exp\left(\beta \sum_{a=1}^n
\sigma^a \sigma_2^a \sigma_3^a \right)
\end{equation}
It turns out that there exist three solutions, corresponding to a
paramagnetic, a ferromagnetic and a spin-glass phase of the system.
 
The simplest solution is the {\it paramagnetic} one,
$c_{pm}(\vec\sigma)=1/2^n$ for all $\vec\sigma$. This solution is
expected to be valid at sufficiently high temperature.  The
paramagnetic free-energy density is given by
\begin{equation}
  \label{eq:fpm}
-\beta f_{pm} = \ln 2 + {k\!+\!1 \over 3}  \ln \cosh(\beta)\ .
\end{equation}

At lower temperature there appears a {\it ferromagnetic}
solution. There, replica symmetry (RS) is expected to hold, and
$c(\vec\sigma)$ depends on $\vec\sigma$ only via $\sum_a\sigma^a$.
Furthermore there are no site to site fluctuation since all nodes of
the hyper-tree are equivalent, so one can look for a ferromagnetic
solution of the type:
\begin{equation}
c(\vec\sigma)=\frac{e^{\beta h\sum_a\sigma^a}}{(2\cosh\beta h)^n} \ .
\label{eq:rs}
\end{equation}
This is indeed a stationary solution provided the effective magnetic
field $h$ is a solution of the equation:
\begin{equation}
\tanh(\beta h)=\tanh(\beta) \tanh(\beta k h)^2
\end{equation}
Nontrivial solutions (with $h \ne 0$) exist only for temperatures
below a certain $T_{ms}$, and disappear abruptly above $T_{ms}$.  The
resulting free energy reads:
\begin{equation}
-\beta f_{fm} = \ln 2 + {k\!+\!1 \over 3} \ln\left[
\cosh(\beta)\cosh(\beta k h)^3+\sinh(\beta)\sinh(\beta k h)^3\right]
-k \ln\left[\cosh(\beta (k\!+\!1) h)\right]
\label{eq:ffm}
\end{equation}
The ferromagnetic transition is of first order: one needs to find the
largest, i.e. locally stable, of the two nontrivial solutions for
$h$. For temperatures slightly below $T_{ms}$ the free energy of this
ferromagnetic solution is larger than that of the paramagnetic one and
the latter stays globally stable. Only at a temperature
$T_{fm}<T_{ms}$, the ferromagnetic solution becomes thermodynamically
dominant (lower free energy), inducing a first order transition. For
the case $k=3$, one gets $ T_{ms}=1.63$ and $ T_{fm}=1.21$.

A third solution of (\ref{eq:csigma}) corresponds to a {\it glass}
phase. As we shall see, this phase is easily accessed by Monte Carlo
simulations with decreasing temperature. Its physical origin lies in
the fact that if one of the spins on a plaquette is down, then the
effective interaction among the other two spins in the same plaquette
becomes antiferromagnetic. Therefore there is no restoring force
towards the ferromagnetically ordered state. In order to find the
glass phase in the replica approach, one has to look for a
$c(\vec\sigma)$ which breaks replica symmetry.

We have found a one step replica-symmetry breaking (RSB) solution,
which is even with respect to separate changes of sign of all its
variables. This solution is somewhat involved because we are dealing
with a finite connectivity system.  We need to build up, on each site
$i$, a distribution of local fields $P_i(h)$. The local fields on site
$i$ in the various pure states $\alpha$ are iid variables chosen from
the distribution $P_i(h)$.  This distribution fluctuates from site to
site and the order parameter is a functional: the distribution of the
functions $P_i$. In replica language, this one-step RSB amounts to a
solution $ c_{rsb}(\vec\sigma)$ of the saddle point equations with the
following structure:
\begin{equation}
\label{eq:Prsb} c_{rsb}(\vec\sigma) = \int d\lambda\, \mu(\lambda)
\prod_{a=1}^{n/m} \left[\int du_a \phi(u_a|\lambda) \exp \left(\beta
u_a \sum_{b=(a-1)m+1}^{am}\sigma^b\right) \right] \ .
\end{equation}
It depends on the real (for $n\to 0$) number $m$, on the function
$\mu$ which is a probability distribution, and on the functions
$\phi(u|\lambda)(2\cosh(\beta u))^m$ which are probability
distributions on $u$ conditioned to a given value of $\lambda$. The
link between the replica approach and the distributions $P_i(h)$ is
reviewed in~\cite{MePa}.

Despite difficulties in finding the saddle point, it is easy to see
that the spin-glass solution is stable against ferromagnetic
fluctuations.  Indeed, the explicit computation of the replica Hessian
matrix corresponding to free energy (\ref{eq:fn}) shows that if
$c(\vec{\sigma})$ is even, then the matrix is positive definite on the
subspace of functions $v(\vec{\sigma})$ which are odd in at least one
variable\footnote{The result holds more generally for diluted p-spin
models for $p\ge 3$, while for $p=2$, the spin-glass solution exists,
but is unstable against ferromagnetic fluctuations.}.

\begin{figure}
\twofigures[width=0.5\textwidth]{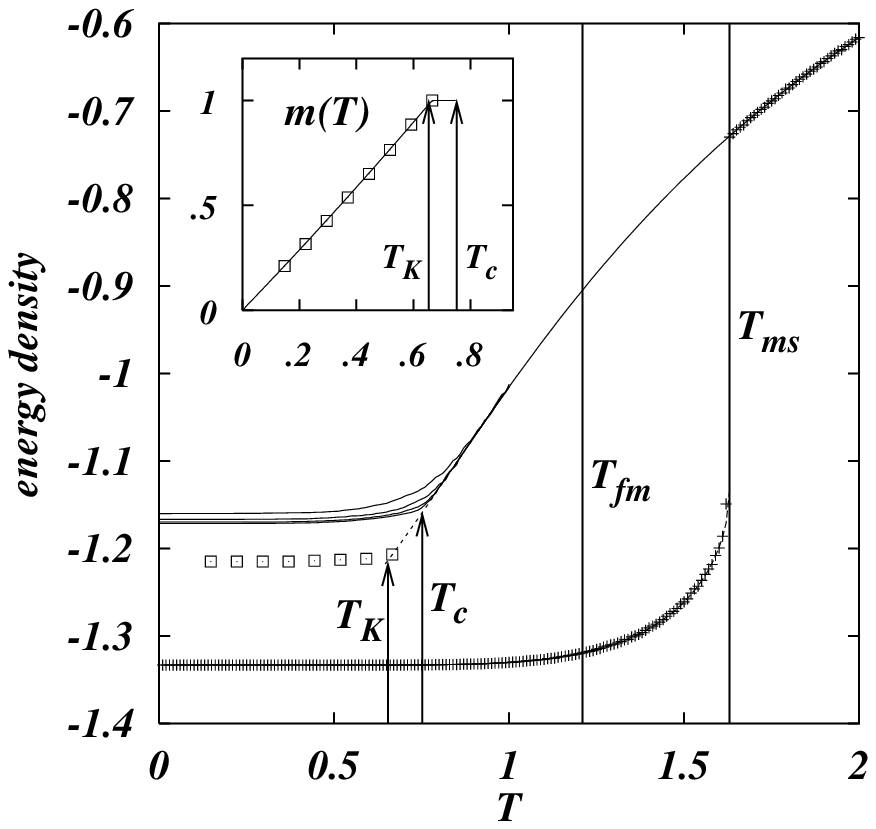}{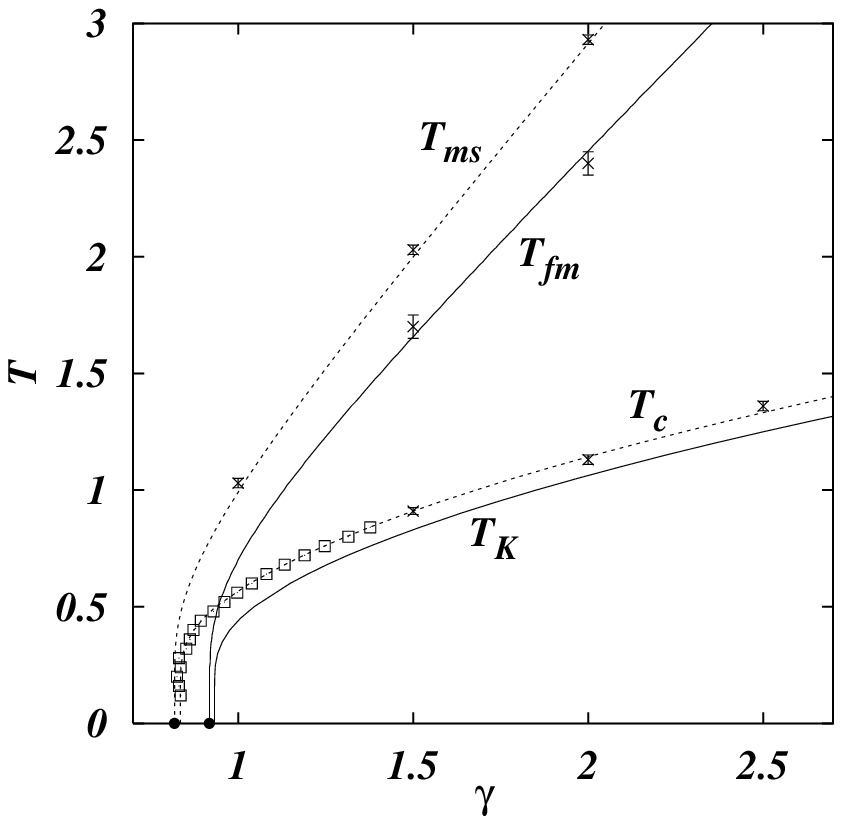}
\caption{Average energy as a function of the temperature for the
ferromagnetic model with $k=3$.  The rightmost vertical lines are the
analytic prediction for the ferromagnetic transitions, while the
leftmost vertical arrows are those for the spin glass transitions.
Continuous lines: results of simulated annealing experiments (cooling
rates from $10^2$ to $10^5$ MCS per $\Delta T=0.01$), where the
magnetization stay zero and the energy converges to the spin glass
threshold energy.  Squares: static energy in the glassy phase.
Crosses: the system is initialized fully magnetized and then slowly
heated.  Data are indistinguishable from the superimposed analytic
curve.  INSET: saddle point value for $m$ as a function of
temperature.  The line is the fit $1.41 T+0.138 T^2$ to $T<T_K$ data.}
\label{cool_heat}
\caption{Continuous (dashed) lines represent the static (dynamic)
transition lines for a 3-spin model with average connectivity
$3\gamma$ calculated with a variational Ansatz.  Upper (lower) lines
refer to the ferromagnetic (spin glass) transition.  Crosses (with
errors): estimations of the critical lines from Monte Carlo
simulations.  Squares: results of the algorithm which gives the exact
1-RSB solution.  Black dots on the $T=0$ axis mark the exact results
for the ferromagnetic model.}
\label{crit_lines}
\end{figure}

We have used the method recently proposed in~\cite{MePa} which allows
to determine the saddle point by a population dynamics of local
fields.  We have found that a nontrivial solution appears for $m=1$
below a certain dynamical temperature $T_c$. This solution appears
discontinuously (the fields are not small close to the transition).
Below a temperature $T_K$ this solution becomes thermodynamically
relevant (letting aside the ferromagnetic state) and the saddle point
value of $m$ is smaller than one (see inset of Fig.~\ref{cool_heat}).
Between $T_K$ and $T_c$ the parameter $m$ sticks to one and the
equilibrium free energy is that of the paramagnetic phase. This
intermediate phase corresponds, as usually, to broken ergodicity with
exponentially many ergodic components, i.e. extensive configurational
entropy.  For the case $k=3$ we have found $T_c=0.745(5)$ and
$T_K=0.660(5)$, while the configurational entropy at $T_c$ equals
$S_{Conf}(T_c)=0.063(5)$ and, as it should, vanishes at $T_K$.

A simple analytic approximation to this one-step RSB result can be
obtained by a variational approximation. In replica language, one can
use for instance an approximate form for the order parameter
$c(\vec\sigma)$ of the type~\cite{BiMoWe}
\begin{equation}
c_{rsb}^{(var)}(\vec\sigma) = \prod_{a=1}^{n/m} \frac{\int D u
\exp(\beta \Delta u \sum_{b=(a-1)m+1}^{am}\sigma^b) }{\int D u (2
\cosh \beta \Delta u)^m}
\label{eq:Prsb_var}
\end{equation}
where $D u=du\exp(-u^2/2)/\sqrt{2\pi}$.  The free-energy density
resulting from this Ansatz has to be optimized with respect to the
variational parameters $\Delta$ and $m$. We find $T_c^{var}=0.752$ and
$T_K^{var}=0.654$, which coincides well with the numerical solution of
the exact saddle-point equation.
 
The temperature $T_c$ is a dynamical temperature where the relaxation
times diverges and the system becomes non ergodic, while the
temperature $T_K$ is the Kauzmann temperature where the system has a
thermodynamic phase transition (as in many other cases the fact that
$T_c$ is different from $T_K$ is made possible by the mean field
nature of the problem~\cite{BouCu}).  This spin glass solution
presents all the properties of the discontinuous spin glasses in which
one-step RSB is exact~\cite{BouCu}. However, we can not exclude
a second static transition with full replica symmetry breaking at a
temperature even smaller then $T_K$, as it happens in the fully
connected Ising $p$-spin model~\cite{gardner}.

Note also, that the RSB solution for $c(\vec\sigma)$ is even under
reversal of all $n$ spins, and, thus, is also valid for the spin-glass
version of the model, where couplings are set to $\pm1$ randomly. The
glassy and the paramagnetic phase turn out to be indistinguishable in
both models.

In order to clarify the physical implications resulting from the
existence of three saddle point solutions, we have performed Monte
Carlo (MC) simulations. First of all we have determined $T_{ms}$ as
the maximum temperature at which the ferromagnetic state is locally
stable.  In Fig.~\ref{cool_heat} the heating curves show the results
of MC simulations -- started a $T=0$ with all spins up -- during which
the temperature is slowly increased.  The static transition
temperature $T_{fm}$ is determined as the point where the time spent
by the simulation in the paramagnetic and in the ferromagnetic state
are equal~\cite{p-spin-3d}.

If we start, however, from a random configuration, the evolution of
large systems ($N=99999$ in our simulations) is completely insensitive
to the presence of a ferromagnetic ground state, the time to find it
being exponentially large in the system size~\cite{RiWeZe}. The system
remains unmagnetized even if quenched below $T_{fm}$, and may thus
undergo only the glass transition. We have measured the stationary
spin-spin correlation function, and have determined the critical point
$T_c$ from the divergence of the correlation time.  The result
$T_c=0.75(1)$ is perfectly compatible with our analytical
estimates. In this respect, the 3-spin ferromagnetic model with finite
connectivity is completely different from any 2-spin unfrustrated
model or any p-spin unfrustrated fully connected one.  Indeed, in the
latter cases the stability argument presented above does not hold and
a simple coarsening dynamics pushes the system towards the
unfrustrated ground state~\footnote{The same trivial dynamics applies
to any unfrustrated model whose connectivity diverges with $N$,
because thermal fluctuations in the magnetization are amplified, until
the ground state is reached.}, thus ruling out the glassy
off-equilibrium behavior discussed above or
in~\cite{p-spin-3d,p-spin-2d}.

In Fig.~\ref{cool_heat} we show the results of a set of cooling
experiments on the unfrustrated model (identical curves have been
obtained for the frustrated model).  When the simulation goes through
the ferromagnetic critical points (the 2 rightmost vertical lines) the
system's evolution is completely unaffected, and the average
magnetization stays near zero. The relaxation process is in fact
strongly slowed down only when the spin glass critical points are
reached (marked by the 2 leftmost vertical arrows).  Note also that
below the spin glass critical point the energy relaxation is almost
absent and the asymptotic energy strongly depends on the cooling rate.
Indeed the most effective relaxation is the one happening close to the
critical point. All the above features are typical for structural
glasses.

We have checked that some amount of disorder does not destroy the
above picture, by studying a system with fluctuating connectivity, in
which the set $E$ of hyper-edges contains $\gamma N$ randomly chosen
triples $[i,j,k],\ i<j<k$.  In this case the average connectivity
$c=3\gamma$ can be varied continuously and critical lines in the
$(\gamma,T)$ plane can be evaluated.  Fig.~\ref{crit_lines} summarizes
our analytical and numerical findings on this model, which are in full
qualitative agreement with those obtained in the case of a hyper-tree
with fixed connectivity.

The analogy with supercooled liquids can be pushed forward once we
identify the ferromagnetic ground state with the crystalline phase,
the paramagnetic state with the liquid phase and the spin glass state
with the glassy phase. If we do not force the system towards the
ferromagnetic ground state, below $T_{fm}$ the system remains in the
paramagnetic phase and it is thus supercooled.  Decreasing further the
temperature the system undergoes a glass transition, even if no
quenched frustration is present.  The frustration is self-induced by
the dynamics: being unable to find the ferromagnetic ground state, the
system is typically in a configuration where a finite fraction of the
interactions are unsatisfied and no long range ferromagnetic order
arises.  Indeed, the ground state can not be found (in polynomial
time) by simply exploiting local information, the only one available
when a spin is updated in a single spin flip dynamics.  We expect a
long lived glassy regime to be present also in short range versions of
our model, where it should be destabilized by slow enucleation
processes, as already seen in numerical simulation of some three
dimensional systems~\cite{p-spin-3d}.

Spin glasses were originally understood as intrinsically disordered
and frustrated systems. A few years ago it was found that the disorder
is not really necessary to induce a spin glass phase~\cite{berna}.  We
have shown here, studying diluted mean-field models, that quenched
frustration is not necessary either.  Our findings give a strong
support to the use of disordered and frustrated models (e.g. p-spin
glasses) to describe structural glasses, which are by constitution
neither disordered nor frustrated.

\acknowledgments

M.W. acknowledges the hospitality of the ICTP and S.F. that of the
LPTMS. This research has been supported in part
by the SPHINX project of the  European Science Foundation.

\end{document}